\documentstyle[12pt,psfig]{article}

\begin{document}

\begin{titlepage}
\begin{flushright}

NTUA--86--2000

hep-th/0001183 \\

\end{flushright}

\begin{centering}
\vspace{.41in}
{\large {\bf Type 0 Brane Inflation from Mirage Cosmology}}\\

\vspace{.2in}

 {\bf E.~Papantonopoulos}$^{a}$ and {\bf I.~Pappa} $^{b}$ \\
\vspace{.2in}

 National Technical University of Athens, Physics
Department, Zografou Campus, GR 157 80, Athens, Greece. \\

\vspace{1.0in}

{\bf Abstract} \\ \vspace{.1in}

We consider a three-dimensional brane-universe moving in a Type 0
String background. The motion induces on the brane a cosmological
evolution which, for some range of the parameters, exhibits an
inflationary phase.

\end{centering}

\vspace{2.5in}
\begin{flushleft}

$^{a}$ e-mail address:lpapa@central.ntua.gr \\$ ^{b}$ e-mail
address:gpappa@central.ntua.gr

\end{flushleft}

\end{titlepage}

\section{Introduction}

 The proposal that our observable four-dimensional universe is a
 domain wall embedded in a higher dimensional space \cite{Reg}, is intensively
 investigated in various contexts. In an earlier
 speculation, motivated by the long standing hierarchy problem, it
 was proposed \cite{Dim} that the fundamental Planck scale could
 be close to the gauge unification scale, at the price of "large"
 spatial dimensions, the introduction of which explains the
 observed weakness of gravity at long distances. In a similar
 scenario \cite{Rand}, our observed world is embedded in a
 five-dimensional bulk, which is strongly curved. This allows the
 extra dimension not to be very large, and we can perceive
 gravity as effectively four-dimensional.

 This idea of a brane-universe can  naturally be applied to string
 theory. In this context, the Standard Model gauge bosons as well as
 charged matter arise as fluctuations of the D-branes. The universe
 is living on a collection of coincident branes, while gravity and
 other universal interactions is living in the bulk space
 \cite{Pol}. For example, the strongly coupled $E_{8} \times E_{8}$
 heterotic string theory is believed to be an eleven-dimensional
 theory, the field theory limit of which was described by Horava
 and Witten \cite{Wit}. The spacetime manifold includes a compact
 dimension with an orbifold structure. Matter is confined on the
 two ten-dimensional hypersurfaces (nine-branes) which can be seen
 as forming the boundaries of this spacetime.

 For all such theories, an essential issue concerns the
 cosmological evolution of our universe. In the literature, there
 are a lot of cosmological models associated to brane universe
 \cite{Cosm}-\cite{Keh}. In these models one can recognize two
  main approaches.
 In the first one, the domain walls (branes) are static solutions of
 the underlying theory, and the cosmological evolution of our
 universe is due to the time evolution of energy density on the
 domain wall (brane) \cite{Stat}. In the second approach, the
 cosmological evolution of our universe is due to the motion of
 our brane-world in the background gravitational field of the
 bulk \cite{Cha,Keh}.

 In \cite{Cha} the motion of a domain wall (brane) in a higher
 dimensional spacetime was studied. The Israel matching conditions
 were used to relate the bulk to the domain wall (brane) metric, and
 some interesting cosmological solutions were found. In
 \cite{Keh} a universe three-brane is considered in motion
 in ten-dimensional space in the presence of a gravitational field
 of other branes. It was shown that this motion in ambient space
 induces cosmological expansion (or contraction) on our universe,
 simulating various kinds of matter.

 In this work we will examine the motion of a three-brane in a
 background of the type 0 string theory. Type 0 string theories \cite{Tset}
 are interesting because of their connection \cite{Typ0} to four-dimensional
 $SU(N)$ gauge theory. The type 0 string does not have spacetime
 supersymmetry and because of that it contains in its spectrum a
 non-vanishing tachyon field. In spite of the presence of the
 tachyon field these theories in an AdS background are
 stable \cite{Kleb}.

 Employing the technics of ref. \cite{Keh}, we will show that a
 three-brane as it moves in the type 0 string bulk, it inflates.
 Stability of the theory requires that the tachyon field takes a
 constant value \cite{Tset}, and taking also a constant value for the
 dilaton field, the
 theory can be solved exactly in an $AdS \times S^{5}$ background.
 We will show that this
 metric induces on the brane a cosmological evolution which for
 some range of the parameters has an inflationary epoch. As we will
 discuss in the following, the tachyon function $f(T)$ which couples to
 the $F_{5}$ form of the type 0 strings, is crucial for the
 inflationary evolution of the brane-universe.

 Our work is organized as follows. In section two, we briefly
 review the technics of ref. \cite{Keh}. In section three, we discuss
 the type 0 string and for constant dilaton and tachyon fields we
 give an exact solution in an $AdS \times S^{5}$ background. In section
 four, using this solution we find the cosmological evolution of the
 three-brane in a background of type 0 string. Finally in the last
 section  we discuss our results.

 \section{Brane-Universe}

 In \cite{Keh} it was  considered a D-brane moving in a generic
 static, spherically symmetric background. As the brane moves in a
 geodesic, the induced world-volume metric becomes a function of
 time, so there is a cosmological evolution from the brane point
 of view. The metric of a D3-brane is parametrized as

\begin{equation}\label{in.met}
ds^{2}_{10}=g_{00}(r)dt^{2}+g(r)(d\vec{x})^{2}+
  g_{rr}(r)dr^{2}+g_{S}(r)d\Omega_{5}
\end{equation}
 and there is also a dilaton field $\Phi$ as well as a $RR$
 background~$C(r)=C_{0...3}(r)$ with a self-dual field strength. The
 action on the brane is given by

\begin{eqnarray}\label{B.I. action}
  S&=&T_{3}~\int~d^{4}\xi
  e^{-\Phi}\sqrt{-det(\hat{G}_{\alpha\beta}+(2\pi\alpha')F_{\alpha\beta}-
  B_{\alpha\beta})}
   \nonumber \\&&
  +T_{3}~\int~d^{4}\xi\hat{C}_{4}+anomaly~terms
\end{eqnarray}
 The induced metric on the brane is
\begin{equation}\label{ind.metric}
  \hat{G}_{\alpha\beta}=G_{\mu\nu}\frac{\partial x^{\mu}\partial x^{\nu}}
  {\partial\xi^{\alpha}\partial\xi^{\beta}}
\end{equation}
 with similar expressions for $F_{\alpha\beta}$ and
 $B_{\alpha\beta}$.In the static gauge,
 using (\ref{ind.metric}) we can calculate the bosonic part of the
 brane Lagrangian which reads

\begin{equation}\label{brane Lagr}
L=\sqrt{A(r)-B(r)\dot{r}^{2}-D(r)h_{ij}\dot{\varphi}^{i}\dot{\varphi}^{j}}
-C(r)
\end{equation}
where $h_{ij}d \varphi ^{i} d \varphi^{j}$ is the line
 element of the unit five-sphere,and

\begin{equation}\label{met.fun}
  A(r)=g^{3}(r)|g_{00}(r)|e^{-2\Phi},
  B(r)=g^{3}(r)g_{rr}(r)e^{-2\Phi},
  D(r)=g^{3}(r)g_{S}(r)e^{-2\Phi}
\end{equation}

 Demanding conservation of energy $E$ and of total angular
 momentum $ \ell ^{2} $ on the brane, the induced four-dimensional metric
 on the brane is

\begin{equation}\label{fmet}
d\hat{s}^{2}=(g_{00}+g_{rr}\dot{r}^{2}+g_{S}h_{ij}\dot{\varphi}^{i}\dot{\varphi}^{j})dt^{2}
+g(d\vec{x})^{2}
\end{equation}
 with

\begin{equation}\label{functions}
\dot{r}^{2}=\frac{A}{B}(1-\frac{A}{(C+E)^{2}}\frac{D+\ell^{2}}{D}),
h_{ij}\dot{\varphi}^{i}\dot{\varphi}^{j}=\frac{A^{2}\ell^{2}}{D^{2}(C+E)^{2}}
\end{equation}
 Using (\ref{functions}), the induced metric becomes

\begin{equation}\label{fin.ind.metric}
d\hat{s}^{2}=-d\eta^{2}+g(r(\eta))(d\vec{x})^{2}
\end{equation}
 with $\eta$ the cosmic time which is defined  by
\begin{equation}\label{cosmic}
 d\eta=\frac{|g_{00}|g^{\frac{3}{2}}e^{-\Phi}}{|C+E|}dt
\end{equation}

 This equation is the standard form of a flat expanding universe.
If we define the scale factor as $\alpha^{2}=g$ then we can
calculate the Hubble constant $H=\frac{\dot{\alpha}}{\alpha}$,
where dot stands for derivative with respect to cosmic time. Then
we can interpret the quantity $(\frac{\dot{\alpha}}{\alpha})^{2}$
as an effective matter density on the brane with the result
\begin{equation}\label{dens}
\frac{8\pi}{3}\rho_{eff}=\frac{(C+E)^{2}g_{S}e^{2\Phi}-|g_{00}|(g_{S}g^{3}+\ell^{2}e^{2\Phi})}
{4|g_{00}|g_{rr}g_{S}g^{3}}(\frac{g'}{g})^{2}
\end{equation}

Therefore the motion of a D3-brane on a general spherically
symmetric background had induced on the brane a matter density. As
it is obvious from the above relation, the specific form of the
background will determine the cosmological evolution on the brane.

\section{Type 0 string background}

Type 0 string theory is interesting because except its connection
to gauge theories it contains a tachyon field. Tachyonic fields in
ordinary field theory create instabilities. In cosmology on the
contrary, the time evolution of a tachyon field plays an important
$r\hat{o}le$. In two dimensions because the tachyon field is a
matter field has important consequences in cosmology \cite{Dia},
and it can give a solution to the "gracefull exit" problem
\cite{Pap}. In four dimensions its effect to cosmology has been
examined by various authors \cite{Perry}. We will show that in the
type 0 string the tachyon field can induce inflation on the brane.
The action of the type 0 string is given by \cite{Tset}
\begin{eqnarray}\label{action}
S_{10}&=&~\int~d^{10}x\sqrt{-g}\Big{[} e^{-\Phi} \Big{(}
 R+(\partial_{\mu}\Phi)^{2} -\frac{1}{4}(\partial_{\mu}T)^{2}
-\frac{1}{4}m^{2}T^{2}-\frac{1}{12}H_{\mu\nu\rho}H^{\mu\nu\rho}\Big{)}
\nonumber \\&& - \frac{1}{2}(1+T+\frac{T^{2}}{2})|F_{5}|^{2}
\Big{]}
\end{eqnarray}

The equations of motion which result from this action are
\begin{equation}\label{dilaton}
  2\nabla^{2}\Phi-4(\nabla_{n}\Phi)^{2}-\frac{1}{2}m^{2}T^{2}=0
\end{equation}

\begin{eqnarray}\label{metric}
  &R_{mn}&+2\nabla_{m}\nabla_{n}\Phi-\frac{1}{4}\nabla_{m}T\nabla_{n}T
  -\frac{1}{4\cdot4!}e^{2\Phi}f(T) \Big{(}F_{mklpq}F_{n}~^{klpq}\nonumber
  \\&&  - \frac{1}{10}G_{mn}F_{sklpq}F^{sklpq} \Big{)}=0
\end{eqnarray}

\begin{equation}\label{Tachyon}
  (-\nabla^{2}+2\nabla^{n}\Phi\nabla_{n}+m^{2})T
  +\frac{1}{2\cdot5!}e^{2\Phi}f'(T)F_{sklpq}F^{sklpq}=0
\end{equation}

\begin{equation}\label{F}
  \nabla_{m} \Big{(}f(T)F^{mnkpq} \Big{)}=0
\end{equation}
The tachyon is coupled to the $RR$ field through the function
\begin{equation}\label{ftac}
 f(T)=1+T+\frac{1}{2} T^{2}
\end{equation}
In the background where the tachyon field acquires vacuum
expectation value $T_{vac}=-1$, the tachyon function (\ref{ftac})
takes the value $f(T_{vac})=\frac{1}{2}$ which guarantee the
stability of the theory \cite{Tset}.

The equations (\ref{dilaton})-(\ref{F}) can be solved using the
following ansatz for the metric

\begin{eqnarray}
ds^{2}_{10}=g_{00}(r)dt^{2}+g(r)(d\vec{x})^{2}+
  g_{rr}(r)dr^{2}+g_{S}(r)d\Omega_{5} \nonumber
\end{eqnarray}
Moreover one can find the electrically charged three-brane if
 the following
ansatz for the $RR$ field

\begin{equation}\label{Form}
  C_{0123}=A(r),   F_{0123r}=A'(r)
\end{equation}
and a constant value for the dilaton field $\Phi=\Phi_{0}$ is used

\begin{equation}\label{Sol}
  g_{00}=-H^{-\frac{1}{2}},
  g(r)=H^{-\frac{1}{2}},  g_{S}(r)=H^{\frac{1}{2}}r^{2},
g_{rr}(r)=H^{\frac{1}{2}},    H=1+\frac{e^{\Phi_{0}}Q}{2r^{4}}
\end{equation}

\section{Brane-inflation}

We consider a D3-brane moving along a geodesic in the background
of a type 0 string. The induced metric on the brane (\ref{fmet})
using the background solution (\ref{Sol}) is
\begin{equation}\label{Ind.Sol}
 d\hat{s}^{2}=(-H^{-\frac{1}{2}}+H^{\frac{1}{2}}\dot{r}^{2}
 +H^{\frac{1}{2}}r^{2}h_{ij}\dot{\varphi}^{i}\dot{\varphi}^{j})dt^{2}
 +H^{-\frac{1}{2}}(d\vec{x})^{2}
\end{equation}
From eq.(\ref{F}) the $RR$ field $C=C_{0123}$ using the ansatz
(\ref{Form}) becomes
\begin{equation}\label{cbar}
  C^{~'}=2 Q g^{2}g^ {-\frac{5}{2}}_{s}\sqrt{g_{rr}}f^{-1}(T)
\end{equation}
where Q is a constant.Using again the solution (\ref{Sol}) the
$RR$ field can be integrated to give

\begin{equation}\label{Cterm}
C=e^{-\Phi_{0}}f^{-1}(T)(1+\frac{e^{\Phi_{0}}Q}{2r^{4}})^{-1}+Q_{1}
\end{equation}
where $Q_{1}$ is a constant. The effective density on the brane
(\ref{dens}), using eq.(\ref{Sol}) and (\ref{cbar})  becomes
\begin{equation}\label{cre}
\frac{8\pi}{3}\rho_{eff}=\frac{1}{4}[(f^{-1}(T)+EHe^{\Phi_{0}})^{2}-(1+\frac{\ell^{2}e^{
2\Phi_{0}}}{2}H)]
\frac{Q^{2}e^{2\Phi_{0}}}{r^{10}}H^{-\frac{5}{2}}
\end{equation}
where the constant $Q_{1}$ was absorbed in a redefinition of the
parameter $E$. Identifying $g=\alpha^{2}$ and using
$g=H^{-\frac{1}{2}}$ we get from (\ref{cre})
\begin{eqnarray}\label{aro}
\frac{8\pi}{3}\rho_{eff}&=&(\frac{2e^{-\Phi_{0}}}{Q})^{\frac{1}{2}}
 \Big{[} \Big{(} f^{-1}(T)+\frac{Ee^{\Phi_{0}}}{\alpha^{4}} \Big{)}^{2}
-\Big{(}1+\frac{\ell^{2}e^{2\Phi_{0}}}
{\alpha^{6}}(\frac{2e^{-\Phi_{0}}}{Q})^{\frac{1}{2}}\nonumber \\&&
(1-\alpha^{4})^{\frac{1}{2}} \Big{)}  \Big{]} (1-\alpha^{4})
^{\frac{5}{2}}
\end{eqnarray}
From the relation $g=H^{-\frac{1}{2}}$ we find
\begin{equation}\label{ro}
  r= (\frac{\alpha^{4}}{1-\alpha^{4}})
  ^{\frac{1}{4}}(\frac{Qe^{\Phi_{0}}}{2})^{\frac{1}{4}}
\end{equation}
This relation restricts the range of $\alpha$ to $0\leq \alpha
<1$, while the range of $r$ is $0\leq r< \infty$. We can calculate
the scalar curvature of the four-dimensional universe as
\begin{equation}\label{curv}
  R_{brane}=8\pi(4+\alpha\partial_{\alpha})\rho_{eff}
\end{equation}
If we use the effective density of (\ref{aro}) it is easy to see
that $R_{brane}$ of (\ref{curv}) blows up at $\alpha=0$. On the
contrary if $r\rightarrow 0$,then the $ds^{2}_{10}$ of
(\ref{in.met}) becomes
\begin{equation}\label{ads}
ds^{2}_{10}= \frac{r^{2}}{L} (-dt^{2}+(d\vec{x})^{2})+
      \frac{L}{r^{2}} dr^{2}+  L d\Omega_{5}
\end{equation}
with $L=(\frac{e^{\Phi_{0}}Q}{2})^{\frac{1}{2}}$.This space  is a
regular $AdS \times S^{5}$ space.

Therefore the brane develops an initial singularity as it reaches
$r=0$, which is a coordinate singularity and otherwise a regular
point of the $AdS_{5}$ space. This is another example in Mirage
Cosmology \cite{Keh} where we can understand the initial
singularity as the point where the description of our theory
breaks down.

If we take $\ell^{2}=0$, set the function $f(T)$ to each minimum
value and also taking $\Phi_{0}=0$, the effective density
(\ref{aro}) becomes
\begin{equation}\label{laro}
\frac{8\pi}{3}\rho_{eff}=(\frac{2}{Q})^{\frac{1}{2}}
\Big{(}(2+\frac{E}{\alpha^{4}})^{2} -1 \Big{)} (1-\alpha^{4}) ^{\frac{5}{2}}
\end{equation}
As we can see in the above relation, there is a constant term,
coming from the tachyon function $f(T)$. For small $\alpha$ and
for some range of the parameters $E$ and $Q$ it gives an
inflationary phase to the brane cosmological evolution. In Fig. 1
we have plotted $\rho_{eff}$ as a function of $\alpha$ for $Q=2$.
 Note here that $E$ is constrained from
(\ref{functions}) as $C+E\geq0$. In our case using (\ref{Cterm})
we get $E\geq -2\alpha^{4}$, therefore $E$ can be as small as we
want.


\begin{figure}[h]
\centering
\centerline{\hbox{\psfig{figure=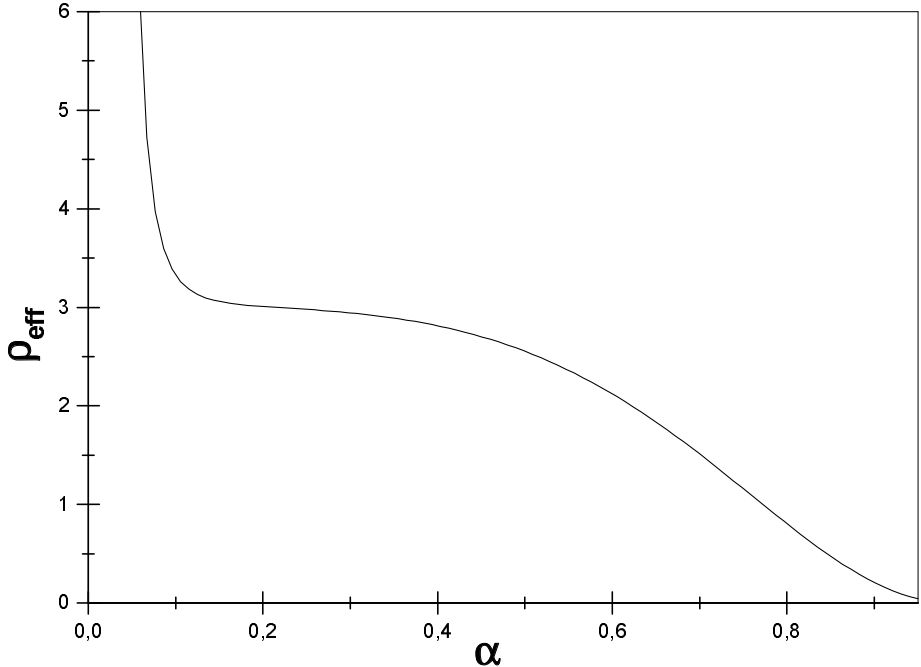,height=10cm}}} \caption
{The induced energy density on the brane as a function of the
brane scale factor.}
\end{figure}

As the brane moves away from $r=0$ to larger values of $r$, the
universe after the inflationary phase enters a radiation dominated
epoch because the term $\alpha^{-4}$ takes over in (\ref{laro}).
As the cosmic time $\eta$ elapses the $\alpha^{-8}$ term dominates
and finally when the brane is far away from $r=0$, the term which
is controlled by the angular momentum $\ell^{2}$ gives the main
contribution to the effective density. Non zero values of
$\ell^{2}$ will give negative values for $\rho_{eff}$. We expect
that at later cosmic times there will be other fields, like gauge
fields, which will give a different dynamics to the cosmological
evolution and eventually cancel the negative matter density.

\section{Discussion}

We had followed the movement of a probe brane along a geodesic in
the background of type 0 strings. Assuming that the universe is
described by a three-dimensional brane, we found that the movement
of the brane in the background, induces a cosmological evolution
of the universe. As the brane-universe moves along smaller values
of the radial coordinate $r$ it contracts while as $r$ becomes
larger it expands. An observer on the brane will see, after the
initial singularity, the universe to enter an inflationary period
and as the scale factor $\alpha$ grows to larger values, the
universe to be dominated by radiation.
    The background we considered is type 0 string. To get an exact
solution in an $AdS \times S^{5}$ background we consider a vacuum
where the tachyon $T$ and the dilaton $\Phi$ are constant. In this
background the tachyon function $f(T)$ which appears as a coupling
of the tachyon to the $RR$ field is a constant. In our analysis
this constant value gives the inflationary phase of the
brane-universe evolution. If we move away from the minimum of
$f(T)$, and the dilaton field is not a constant, the equations
(\ref{dilaton})-(\ref{F}) cannot be solved exactly. However there
are approximate solutions to the above equations in the infrared
and the ultraviolet \cite{Minah}.
    Work is in progress \cite{Lies} to develop a realistic phenomenological
model for "Mirage Inflation" in which all the astrophysics
constraints will be satisfied. We also study the effect of a non
constant tachyon and dilaton field to the cosmological evolution
of the brane-universe.

\section*{Acknowledgement}
We would like to thank A. Kehagias  for many discussions on the
brane-world.

 \end{document}